\documentclass[twocolumn,aps,pra,groupedaddress,showpacs]{revtex4}
\usepackage{epsfig,amssymb,amsmath}
\def\comment#1{}\def\labell#1{\label{#1}}
\def\cm{{\cal M}}\def\cn{{\cal N}}\def\ms{{\mathbb S}}
\def\vi#1{\stackrel{\mbox{\tiny $\leftrightarrow$}}{#1}}

\def\vu#1{\stackrel{\mbox{\tiny $\leftarrow$}}{#1}}
\begin{document}
\title{Additivity properties of a Gaussian channel}
\author{Vittorio Giovannetti$^1$ and Seth Lloyd$^{1,2}$}
\affiliation{$^1$Massachusetts Institute of Technology
  -- Research Laboratory of Electronics\\$^2$
Massachusetts Institute
  of Technology -- Department of Mechanical Engineering\\ 77
  Massachusetts Ave., Cambridge, MA 02139, USA}

\begin{abstract}
The Amosov-Holevo-Werner conjecture implies the additivity of 
the minimum R\'enyi entropies at the output of a channel. 
The conjecture is proved true for all R\'enyi
entropies of integer order greater than two
 in a class of Gaussian Bosonic channel
where the input signal is randomly displaced 
or where it is coupled linearly to an external environment.
\end{abstract}
 \maketitle 
One of the most challenging open questions of quantum communication
theory is the additivity of the various quantities characterizing
the information transmission in a channel \cite{shorequiv}.
The issue at hand is whether quantum entanglement 
is able to improve the performance
of classical protocols \cite{chuang,shor}. 
The supposed additivity of the Holevo information \cite{hole}
is the most important example of this kind of issue. 
The maximum of this quantity 
over all possible encoding procedures is known 
to provide the capacity $C_1$ in transmitting classical 
information for a single use of the channel.
However, if the sender of the message is 
allowed to encode messages in entangled states among $m$ 
successive uses of the communication line, then the resulting capacity
per channel use might be higher than $C_1$ \cite{HSW}. 
For this reason to compute the 
ultimate classical capacity $C$ of the channel it
is  necessary to introduce a regularization of the 
Holevo information where
a limit $m \rightarrow \infty$ has to be performed \cite{shorequiv,HSW}.
All this could be avoided if only the 
Holevo information was shown to be an additive quantity.
Up to now no channel has been found for
which this regularization is necessary:
on the contrary all the channels for which the value
of $C$ has been calculated have additive
Holevo information~\mbox{\cite{add,king1,c}}. 

The additivity of the Holevo information has been linked 
to the additivity of  other 
relevant quantities in Ref. \cite{shorequiv}. In particular,
it is known that proving the additivity of the 
Holevo information is equivalent to proving the additivity
of the minimum von Neumann entropy $\ms$ at the output of the
channel. 
Given a channel described by the 
completely positive (CP) linear
map $\cal M$ on the input space $\cal H$,
this quantity is defined as
\begin{eqnarray}
\ms({\cal M}^{\otimes m}) \equiv \min_{\rho\in {\cal H}^{\otimes m}} 
S\left({\cal M}^{\otimes m}(\rho)\right)\;, \labell{entropy}
\end{eqnarray}
where the minimization is performed over 
all the possible input states $\rho$ of $m$ successive uses of the channel, 
and where $S(\rho)\equiv -\mbox{Tr}[\rho \ln \rho]$. 
The additivity hypothesis requires $\ms({\cal M}^{\otimes m})$ to be
equal to $m$ times the minimum entropy for a single
use of the channel  $\ms({\cal M})$: this conjecture  seems simpler
to study than the additivity of the Holevo information and
some authors have focused their attention to it
\cite{ruskai1,entro,entro1,king1}.
As a matter of fact the alleged additivity of the $\ms$ is just
a particular instance of the Amosov-Holevo-Werner conjecture
\cite{asomov} which requires the maximum of the output $z$-norm $\nu_z({\cal M}^{\otimes m})$
of the channel to be multiplicative, i.e. 
it requires that for $m$ integer 
\begin{eqnarray}
\nu_z({\cal M}^{\otimes m})\equiv \max_{\rho\in {\cal H}^{\otimes m}} 
\|{\cal M}^{\otimes m} (\rho)\|_z =  \, \left[\nu_{{z}}({\cal M})\right]^m
\labell{normach}
\end{eqnarray}
where the maximization is performed again over all the input states
of $m$ uses of the channel and where 
\begin{eqnarray}
\|A\|_z \equiv \left( \mbox{Tr}|A|^z \right)^{1/z}
\qquad z\geqslant 1 \labell{norme}
\end{eqnarray}
is the $z$-norm of the operator $A$. In other words,
the conjecture requires the maximization in the left-hand-side  of
Eq.~(\ref{normach}) to be achieved on non entangled states of ${\cal
H}^{\otimes m}$.
The connection between the property of Eq.~(\ref{normach}) and
the additivity of the minimal output entropy, can be established through 
the quantum R\'enyi entropy
\begin{eqnarray}
S_{z}(\rho)\equiv-\frac {\ln\mbox{Tr}[\rho^{z}]}{{z}-1}
\;\labell{defrenyi}.
\end{eqnarray}
This quantity is monotonic with respect to the 
${z}$-norm (\ref{norme}) of the state $\rho$. For $z=2$ the R\'enyi entropy is
a function of the linear entropy $1-\mbox{Tr}[\rho^2]$ and
in the limit ${z}\to 1$ it tends to the von Neumann entropy
\cite{zyc}.
As for the case of $S$ one can define the minimal value
\begin{eqnarray}
\ms_{z}(\cm^{\otimes m})\equiv\min_{\rho\in {\cal H}^{\otimes
    m}}\; S_{z}(\cm^{\otimes m}(\rho)).
\;\labell{defmsr}
\end{eqnarray}
If the Amosov-Holevo-Werner conjecture (\ref{normach}) is true then
the minimum output $z$-R\'enyi entropy is additive and {\em vice versa}. 
Moreover, 
if such property is
verified for values of $z$ arbitrarily close to $1$ then the additivity
of $\ms$ (and hence of the Holevo information) 
follows \cite{asomov1}.

In this paper we will analyze the conjecture (\ref{normach}) for a
set Gaussian channels and prove that it
is true for all integer~$z$. The material is
organized as follows. 
In Sect. \ref{s:channel} we introduce the simple Gaussian channel model
${\cal N}_n$ and in Secs. \ref{s:proof} and \ref{s:optimal}
we show that the conjecture (\ref{normach}) applies to this
channel when $z$ is integer.
 In Sec. \ref{s:other} we analyze the case of generic~$z$
giving some bounds for  $\nu_z({\cal N}_n^{\otimes m})$. In
Sec. \ref{s:altri} we generalize the results of the
first section to a whole class of Gaussian channels.

\section{The channel model}\labell{s:channel}
The channel we analyze here  
is a Bosonic linear channel
where the photonic signal from the sender is displaced
randomly by the environment.
This system is described
by the CP map ${\cal N}_n$ which transforms
the input state of the channel into the output
\begin{eqnarray}
{\cal N}_n (\rho)=\int d^2 \mu \;
P_{n}(\mu) \; D(\mu)\rho D^{\dagger}(\mu)
\labell{due}
\end{eqnarray} 
where, for $n\geqslant 0$, $P_n(\mu)$ is the circularly
symmetric probability distribution
\begin{eqnarray}
P_n(\mu)=\frac{e^{-|\mu|^2/n}}{\pi n},
\labell{tre}
\end{eqnarray}
and $D(\mu)\equiv \exp(\mu a^\dag - \mu^* a)$ is the displacement
operator of the annihilation $a$ of the input signal.  
This channel is Gaussian,  i.e. it
maps the set of input states 
with Gaussian symmetrically characteristic function
into itself  \cite{werner}.  
Moreover, the
map (\ref{due})  is unital (i.e. it transforms the identity operator
in to itself) and it is covariant under
displacement or phase transformation \cite{entro}.
When ${\cal N}_n$ acts on a coherent state
$\rho_\alpha\equiv|\alpha\rangle\langle \alpha|$
 the following transformation takes place,
\begin{eqnarray}
\rho_\alpha \rightarrow 
\cn_n(\rho_\alpha)= D(\alpha)\; \tau(n)\; D^{\dag}(\alpha)
\labell{coherent}\;,
\end{eqnarray} 
with
\begin{eqnarray}
\tau(n)\equiv\frac{1}{n+1}\left(\frac{n}{ n+1}
\right)^{a^\dag a}
\labell{vacuum}\;,
\end{eqnarray} 
the thermal state that gives the output of the
channel for a vacuum input \cite{entro}.
The state $\cn_n(\rho_\alpha)$ has $z$-norm~(\ref{norme}) equal to
\begin{eqnarray}
\| \cn_n(\rho_\alpha) \|_z \equiv \left[\frac{1}{(n+1)^z-n^z}\right]^{1/z}
 \labell{normalpha}
\end{eqnarray}
which does not depend on $\alpha$ since it is
invariant under the unitary transformation $D(\alpha)$. 
In Ref. \cite{entro1} the
right-hand-side of Eq.~(\ref{normalpha}) was shown
to coincide with  the $z$-norm of the single use of the
channel $\nu_{{z}}({\cal N}_n)$, at least for all $z=k$ integer.
In Sec. \ref{s:proof} we will generalize this result 
showing that, for all integer $k$, the classical
channel satisfies the identity,
\begin{eqnarray}
\nu_{{k}}({\cal N}_n^{\otimes m}) =  \,  
\left[\frac{1}{(n+1)^k-n^k}\right]^{m/k}
\labell{normach1}
\end{eqnarray}
hence proving the conjecture (\ref{normach}) for integer $z=k$ for the
channel ${\cal N}_n$.
Equations (\ref{normalpha}) and (\ref{normach1}) imply that 
the maximization implicit in the
definition of $\nu_{{k}}({\cal N}_n^{\otimes m})$ is achievable with
separable input states of the form $|\alpha_1\rangle_1 \otimes
\cdots \otimes |\alpha_m\rangle_m$, i.e. by feeding the channel
with a coherent state in each of the $m$
successive uses. This result will be proved explicitly in  
Sec \ref{s:optimal}.

\subsection{The proof}\labell{s:proof}

In this section we show that Eq.~(\ref{normach1}) applies  for
integer~$z$. Clearly, the right-hand-side of this equation
is a lower bound for the left-hand-side:
the former is in fact the output $z$-norm associated 
to the input signal where the $m$ uses
of the channel have been prepared in coherent states.
To prove the equality in  Eq.~(\ref{normach1}) it is hence sufficient to show
that the right-hand-side is also an upper bound for 
$\nu_{{k}}({\cal  N}_n^{\otimes m})$,
i.e. that for all input states $\rho\in{\cal
  H}^{\otimes m}$ the following
inequality applies,
\begin{eqnarray}
\mbox{Tr} \left\{[\cn_n^{\otimes m}(\rho)]^{k} \right\}
\leqslant \left[\frac{1}{(n+1)^z-n^z}\right]^{m} \labell{thesis}
\end{eqnarray}
The method to derive this property
is similar to the one given in
Ref. \cite{entro1} where an analogous approach was used to calculate
the minimum output R\'enyi entropy (\ref{defrenyi})
of integer order for a single
channel use ($m=1$). The only difference is that here
we are dealing with an extra tensorial structure associated with
$m> 1$.
For the sake of clarity we divide the proof in two separate 
parts.
First we show that the quantity on the left-hand-side of
Eq.~(\ref{thesis}) can be expressed as the expectation value
of an diagonalizable \cite{TYPO} operator $\Theta$ which acts on the Hilbert space
$({\cal H}^{\otimes m})^{\otimes k}$: this allows us to derive an upper bound
for $\mbox{Tr} \left\{[\cn_n^{\otimes m}(\rho)]^{k} \right\}$ by
considering the eigenvalue $\lambda_0$ of $\Theta$ with maximum absolute value.
The second part of the proof is devoted to the analysis of the tensorial
structure of $\Theta$ and to the proof that $\lambda_0$ 
coincides with the left-hand-side
 of Eq.~(\ref{thesis}).

\paragraph*{Part one:--}
Without loss of generality we  
can assume the initial state of the
$m$ uses of the channel 
to be pure, i.e. $\rho=|\psi\rangle\langle\psi|$. 
The convexity of the norm 
(\ref{norme}) guaranties in fact 
that the maximization in Eq.~(\ref{normach}) is achievable 
with pure input states 
\cite{asomov,asomov1}: 
thus if Eq.~(\ref{thesis})
holds for all pure states, then it is valid also for all 
the other channel inputs.
In general $|\psi\rangle$ will be entangled among the various
channel uses and the corresponding output state will be 
\begin{eqnarray}
&&\cn_n^{\otimes m}(\rho)=
\int d^2\mu_1\cdots d^2\mu_{m}
P_n(\mu_1)\cdots P_n(\mu_{m})\nonumber\\
&&\quad D_1(\mu_1) \cdots D_m(\mu_m)\; \rho \;
D^\dag(\mu_1) \cdots D_m^\dag(\mu_{m})\;,\labell{rhoout} 
\end{eqnarray}
where $D_{r}(\mu)\equiv 
\exp(\mu a_{r}^\dag - \mu^* a_{r})$ is
the displacement associated with the annihilation operator $a_{r}$ of
the ${r}$th use of the channel.
Equation (\ref{rhoout}) can be expressed in a more compact form
by introducing a vectorial notation, where
$\vec{\mu}\equiv(\mu_1,\cdots,\mu_m)$ is a 
complex vector in ${\mathbb C}^{m}$ and
$\vec{a}\equiv(a_1,\cdots,a_m)$. The output state
becomes thus
\begin{eqnarray}
\cn_n^{\otimes m}(\rho)=\int d^2\vec{\mu} \;\;P_n(\vec{\mu})\;
D(\vec{\mu}) \; \rho \;
D^\dag(\vec{\mu}) \;,
\end{eqnarray}
where 
\begin{equation}
P_n(\vec{\mu}) \equiv \frac{\exp[-|\vec{\mu}|^2/n]}{(\pi n)^m}
\end{equation}
and $D(\vec{\mu})=\exp(\vec{\mu}\cdot \vec{a}^{\dag} -
\vec{a}\cdot\vec{\mu}^{\dag})$ is a multi-mode displacement operator where
the input of the  ${r}$th  use of the channel is displaced by $\mu_{r}$.
Consider now for $z=k$ integer the quantity
\begin{eqnarray}
&&\mbox{Tr} \left\{[\cn_n^{\otimes m}(\rho)]^{k} \right\}=
\int d^2\vec{\mu}_1\cdots d^2\vec{\mu}_{k}
P_n(\vec{\mu}_1)\cdots P_n(\vec{\mu}_{k})\nonumber\\
&&\qquad\times\mbox{Tr}[D(\vec{\mu}_1)\rho D^\dag(\mu_1) \cdots 
D(\vec{\mu}_k)\rho D^\dag(\mu_k)]
\;\labell{rhoalla}.
\end{eqnarray}
Since $\rho=|\psi\rangle\langle \psi|$, 
the trace in the integral can be expressed as
the expectation value of an operator $\Theta$ which acts in a
extended Hilbert space $({\cal H}^{\otimes m})^{\otimes k}$ made of
$k$ copies of the initial one. In fact, from the invariance of the
trace under cyclic permutation we have
\begin{widetext}
\begin{eqnarray}\nonumber
&&\mbox{Tr}[D(\vec{\mu}_1)\rho D^\dag(\vec{\mu}_1)
\cdots D(\vec{\mu}_k) \rho D^\dag(\vec{\mu}_{k})]
=\langle\psi|D^\dag(\vec{\mu}_1)D(\vec{\mu}_2)|\psi\rangle
\langle\psi|D^\dag(\vec{\mu}_2)D(\vec{\mu}_3)|\psi\rangle\cdots
\langle\psi|D^\dag(\vec{\mu}_{k})D(\vec{\mu}_1)|\psi\rangle\\
&&\qquad\qquad=\mbox{Tr}\left\{(\rho\otimes\rho\otimes
\cdots\otimes\rho)\left[D_1^\dag(\vec{\mu}_1)
D_1(\vec{\mu}_2)\otimes D_2^\dag(\vec{\mu}_2)
D_2(\vec{\mu}_3)\otimes
\cdots \otimes D_{k}^\dag(\vec{\mu}_{k})D_{k}(\vec{\mu}_1)\right]\right\}
\;,\labell{do}
\end{eqnarray}
\end{widetext}
where the ${k}$ scalar products in the input Hilbert space ${\cal
  H}^{\otimes m}$ in
the first line were replaced with a single expectation value in
$({\cal H}^{\otimes{m}})^{\otimes k}$ in the second line. 
In Eq.~(\ref{do}) the operator $D_s(\vec{\mu})$ 
represents
the multi-mode displacement
that operates on the $s$th copy of ${\cal H}^{\otimes m}$, i.e.
\begin{eqnarray}
D_s(\vec{\mu})=\exp(\vec{\mu}\cdot \vec{a}_s^{\dag} -
\vec{a}_s\cdot\vec{\mu}^{\dag})
\labell{multid}
\end{eqnarray}
where for $s=1,\cdots k$
\begin{eqnarray}
\vec{a}_s\equiv(a_{s1},a_{s2},\cdots,a_{sm})\;,
\labell{avec}
\end{eqnarray}
are the $m$ annihilation
operators pertaining to the $s$th copy of  ${\cal H}^{\otimes m}$.
With this trick Eq.~(\ref{rhoalla}) can be written as,
\begin{eqnarray}\mbox{Tr} \left\{[\cn_n^{\otimes m}(\rho)]^{k} \right\}
=\mbox{Tr}[(\rho\otimes\cdots\otimes\rho)\; \Theta]
\;\labell{mi},
\end{eqnarray}
where each of the $k$ copies of the state $\rho$ is associated 
to one of the multi-mode annihilation operator $\vec{a}_s$ and where
 $\Theta$ is the operator on $({\cal H}^{\otimes m})^{\otimes
  k}$ given by
\begin{eqnarray}
\Theta&=&\int d^2\vec{\mu}_1 \cdots  d^2\vec{\mu}_k
\; \;P_n(\vec{\mu}_1) \cdots P_n(\vec{\mu}_k)\\
&&\qquad \; \times \; D_1^\dag(\vec{\mu}_1)
D_1(\vec{\mu}_2)\otimes 
\cdots \otimes D_{k}^\dag(\vec{\mu}_{k})D_{k}(\vec{\mu}_1)
\labell{re1}
\;.\nonumber\end{eqnarray}
Equation (\ref{mi}) allows us to derive an upper bound
for the quantity on the left-hand-side by 
considering the eigenvalue $\lambda_0$ 
of $\Theta$ with maximum absolute
value, i.e.
\begin{eqnarray}\mbox{Tr} \left\{[\cn_n^{\otimes m}(\rho)]^{k} \right\}
\leqslant |\lambda_0|
\;\labell{mmi}.
\end{eqnarray}
\paragraph*{Part two:--}To calculate $\lambda_0$ it is useful  to analyze
in details the properties of the operator $\Theta$. As shown in 
App. \ref{s:tensore}, this operator has a very simple tensorial
form with respect to the index $r$. 
In fact Eq.~(\ref{re1}) can be decomposed as
\begin{eqnarray}
\Theta = \bigotimes_{r=1}^m \Theta_r \;,
\labell{tensore}
\end{eqnarray}
where, for $r=1,\cdots,m$, the operator $\Theta_r$
 acts on the modes
associated with the annihilation operators
\begin{eqnarray}
\vu{a}_r\equiv (a_{1r}, a_{2r}, \cdots, a_{kr})\;.
\labell{ava}
\end{eqnarray}
In vectorial notation $\Theta_r$ 
can be expressed as
\begin{eqnarray}
\Theta_r
&\equiv&\int \frac{d^2\vu{\mu}_r}{(\pi n)^{k}}
\:{e^{-\vu\mu_r\cdot
C\cdot\vu\mu_r^\dag+\vu\mu_r\cdot
G^\dag\cdot\vu{a}_r^\dag-\vu{a}_r\cdot G
\cdot\vu\mu_r^\dag}}
\;\labell{re},
\end{eqnarray}
where, as in Eq.~(\ref{ava}),  
$\vu{\mu}_r\equiv(\mu_{1r}, \mu_{2r}, \cdots, \mu_{kr})$
is a $k$-element vector, and
where $G$
and 
\begin{eqnarray}
C\equiv\frac\openone n+\frac{A}{2} \;,
\label{defC}
\end{eqnarray}
 are ${k}\times{k}$
real matrices ($\openone$ is the identity).
For $k\geqslant3$, $G$ and $A$ are 
\begin{eqnarray}
G&\equiv&\left[\begin{array}{rrrrrr}
-1&1&0&\cdots&0&0\cr
0&-1&1&\cdots&0&0\cr
0&0&-1&\cdots&0&0\cr
\vdots&&&\ddots&&\cr
0&0&0&\cdots&-1&1\cr
1&0&0&\cdots&0&-1\cr
\end{array}
\right]
\;\labell{defg}
\\
A&\equiv&\left[\begin{array}{rrrrrr}
0&-1&0&\cdots&0&1\cr
1&0&-1&\cdots&0&0\cr
0&1&0&\cdots&0&0\cr
\vdots&&&\ddots&&\cr
0&0&0&\cdots&0&-1\cr
-1&0&0&\cdots&1&0\cr
\end{array}
\right]
\;\labell{defa}.
\end{eqnarray}
For $k=2$ 
$A$ is null,
 while for $k=1$ both $G$ and $A$ are null.
The decomposition~(\ref{tensore}) shows that in the
right-hand-side of Eq.~(\ref{mi}) we have a product of
two operators of  $({\cal H}^{\otimes m})^{\otimes k}$
which have ``orthogonal'' tensorial decomposition: the operator
$\rho\otimes \cdots \otimes \rho$ factorizes with respect
to the index $r=1, \cdots, k$, while $\Theta$ factorizes with
respect to the index $s=1, \cdots, m$ associated with the
successive uses of the channel.
This property is common to all memoryless channels where
the corresponding CP map acts on each channel use independently.
However, the channel model we are considering allows us to
further decompose the operator $\Theta$. In fact,
$G$ and $A$ of Eqs.~(\ref{defg}) and (\ref{defa}) are two  
circulant matrices~\cite{circulant} which commute and possess a common
basis of orthogonal eigenvectors. 
This means that there exists a
$k\times k$ unitary matrix $Y$ for which
\begin{eqnarray}
D&\equiv& Y\:C\:Y^\dag =\openone/n +  Y\:A\:Y^\dag
\nonumber\\
E&\equiv& Y\:G\:Y^\dag \labell{diagonali}
\end{eqnarray}
are diagonal. Since $A$ is antisymmetric, its eigenvalues
$i\xi_j$ are imaginary and the diagonal elements of
$D$ (i.e. the quantities $d_j=1/n + i\xi_j$) have positive real part.
Using these properties we can rewrite the operator $\Theta_r$ of
Eq.~(\ref{re}) in factorized form by performing the change of
integration variables $\vu\nu_r\equiv\vu\mu_r\cdot Y^\dag$ and
introducing the new annihilation operators 
\begin{eqnarray}
\vu b_r\equiv (b_{1r}, b_{2r}, \cdots, b_{kr})
=\vu a_r\cdot Y^\dag
\labell{fra}\;.
\end{eqnarray}
These operations yield
\begin{eqnarray}
\Theta_r=\bigotimes_{j=1}^{k} \Theta_{jr}
\;\labell{fatt},
\end{eqnarray}
with 
\begin{eqnarray}
\Theta_{jr}\equiv\frac 1{n|e_j|^2}\int \frac{d^2\nu}{\pi}
\:{e^{-d_j|\nu|^2/|e_j|^2}}\:D_{b_{jr}}(-\nu)
\;\labell{re2},
\end{eqnarray}
where  $D_{b_{jr}}(\nu)\equiv\exp[\nu b_{jr}^\dag-\nu^*b_{jr}]$ is the
displacement operator associated with $b_{jr}$, while $e_j$ is
the $j$th diagonal elements of the matrix $E$
(i.e. the $j$th eigenvalue of $G$).
As demonstrated in Ref. \cite{entro,caves1}, this expression can be
further simplified, proving that
$\Theta_{jr}$ is diagonal in the Fock
basis of the mode $b_{jr}$ and equal to
\begin{eqnarray}
\Theta_{jr}=\frac{2/n}{2d_j+|e_j|^2}
\left(\frac{2d_j-|e_j|^2}{2d_j+|e_j|^2}\right)^{b_{jr}^\dag
  b_{jr}}
\;,\labell{defqj}
\end{eqnarray}
(for the sake of completeness we give an alternative derivation
of this result in App. \ref{s:ren}).

Equations (\ref{tensore}), (\ref{fatt}) and (\ref{defqj})
show that the eigenvalues of $\Theta$ are products of
eigenvalues of $\Theta_{jr}$.
In particular $\lambda_0$ of Eq.~(\ref{mmi}) is obtained by
taking the eigenvalues of the $\Theta_{jr}$
that have maximum absolute value. 
Since the constants $d_j$ have positive real part,
the quantities we are looking for are $2/[n(2d_j+|e_j|^2)]$,
i.e. they are the eigenvalues of the operators $\Theta_{jr}$ 
associated with the vacuum state 
of the mode $b_{jr}$.
This allows us to express the value of $\lambda_0$ as
\begin{eqnarray}
\lambda_0&=&\prod_{r=1}^m \prod_{j=1}^{k}
\frac{2/n}{2d_j+|e_j|^2}=\left\{\frac{1/n^{k}}
{\det[C+G^\dag G/2]} \right\}^m\nonumber\\&=&
\left[\frac 1{(n+1)^{k}-n^{k}}\right]^m
\;\labell{croma},
\end{eqnarray}
which, replaced in Eq.~(\ref{mmi}), proves the thesis 
(\ref{thesis}). [In deriving the second identity we have used the
invariance of the determinant under the unitary transformation $Y$,
while the last identity can be obtained from the definitions
(\ref{defg}) and (\ref{defa}) by direct calculation of
the determinant itself].

\subsection{Optimal inputs}\labell{s:optimal}

Equations~(\ref{normalpha}) and (\ref{normach1}) 
prove that tensor products of coherent states are optimal since 
they allow
the channel ${\cal N}_n$ to achieve the maximal $k$-norm at the
output of $m$ successive uses.
Here we re-derive this result
by showing that the state 
$\rho \otimes \cdots \otimes \rho$ of Eq.~(\ref{mi})
with $\rho$ given by a tensor product of 
coherent states in the inputs modes $\vec{a}_s$, 
 is an eigenvector of $\Theta$
associated with
the eigenvalue $\lambda_0$ of Eq.~(\ref{croma}).

From the analysis of the previous section we know that
the eigenvectors of $\lambda_0$ can be written as
\begin{eqnarray}
|\Phi\rangle \equiv \bigotimes_{r=1}^m |\Phi_r \rangle_r
\labell{auto}\;,
\end{eqnarray}
where $|\Phi_r\rangle_r$ is an eigenvector of 
$\Theta_r$ of Eq.~(\ref{fatt}) relative to its eigenvalue
with maximum absolute value. For instance, in deriving Eq.~(\ref{croma}) we
have considered the state where each of the 
$\vu{b}_r$ modes is in the vacuum. However this is not the only
possibility. 
In fact,  we
notice that for any ${k}$ the matrices $G$ and $A$ of Eqs.~(\ref{defg})
and (\ref{defa}) have a null
eigenvalue (say for $j=1$), associated with the common normalized 
eigenvector $(1,1,\cdots,1)/\sqrt{k}$. On one hand this means that all the
elements in the first row of the matrix 
$Y$ of Eq.~(\ref{diagonali}) are equal to $1/\sqrt{k}$.
On the other hand, this implies also that
 $e_1=0$, $d_1=1/n$ and,
according to Eq.~(\ref{defqj}),
\begin{eqnarray}
\Theta_{1r}=\openone_{1r}\;,.
\label{unoezero}
\end{eqnarray}
This means that any state of the form
\begin{eqnarray}
|\Phi_r\rangle_r \equiv |\phi_r\rangle_{b_{1r}}\otimes |0\rangle_{b_{2r}}
\otimes \cdots 
\otimes|0\rangle_{b_{kr}} 
\labell{autotopo}\;,
\end{eqnarray}
where the mode $b_{1r}$ is prepared in a generic state $|\phi_r\rangle$
while the other $b_{jr}$ are in the vacuum, is an eigenstate of $\Theta_r$
relative to the eigenvalue with maximum absolute value.
Consider now the case of $|\phi_r\rangle=|\sqrt{k}\alpha_r\rangle$ coherent.
By using the symmetric
characteristic function decomposition \cite{walls} 
we can show that, when expressed in terms of the operators
$\vu{a}_{r}$, the state (\ref{autotopo}) is a tensor product 
of coherent states $|\alpha_r\rangle$. In fact, defining
the complex vector $\vu\gamma\equiv(\sqrt{k}\alpha_r,0,\cdots,0)$ we
can express the state $|\Phi_r\rangle_r$ as,
\begin{eqnarray}
|\Phi_r\rangle_r\langle \Phi_r|&=&\int \frac{d^2\vu\nu}{\pi^{k}}
\:\exp[-|\vu\nu|^2/2 \nonumber \\
&&\qquad+\vu\nu\cdot(\vu b_r-\vu\gamma)^\dag-(\vu b_r-\vu\gamma)
\cdot\vu\nu^\dag]\nonumber\\
&=&\int \frac{d^2\vu\mu}{\pi^{k}}\:\exp[-|\vu\mu|^2/2 \nonumber
\\
&&\qquad 
+\vu\mu\cdot(\vu a_r^\dag-Y^\dag \cdot\vu\gamma)
-(\vu a_r-\vu\gamma\cdot Y)\cdot\vu\mu^\dag]\nonumber\\
&=&|\alpha_r\rangle_{a_{1r}}\langle
\alpha_r|\otimes\cdots\otimes|\alpha_r
\rangle_{a_{kr}}\langle \alpha_r|
\;\labell{vuoto},
\end{eqnarray}
where the second identity is obtained by substituting 
the $\vu\nu$ with $\vu\mu\cdot Y^\dag$ in the integral, while the
third identity derives from the properties of the matrix $Y$
discussed above. The thesis finally follows by replacing this
expression in Eq.~(\ref{auto}),
\begin{eqnarray}
|\Phi\rangle\langle\Phi| &=& \bigotimes_{r=1}^{m} 
(\;|\alpha_r\rangle_{a_{1r}}\langle
\alpha_r|\otimes\cdots\otimes|\alpha_r
\rangle_{a_{kr}}\langle \alpha_r|\;) 
\labell{basta1}\;,
\end{eqnarray}
and observing that this can be represented as $\rho\otimes
\cdots \otimes \rho$ of Eq.~(\ref{mi}) with $\rho=\left( \bigotimes_{r=1}^{m} |
  \alpha_r \rangle\langle \alpha_r | \right)^{\otimes k}$.

\subsection{Upper bound}\labell{s:other}
In this section, starting from the values of the $\nu_z({\cal
  N}_n^{\otimes m})$ for $z$ 
integer derived
in the previous section, we give  some upper bounds for the
channel \mbox{${z}$-norm}  of generic order. 

\begin{figure}[hbt]
\begin{center}
\epsfxsize=.9
\hsize\leavevmode\epsffile{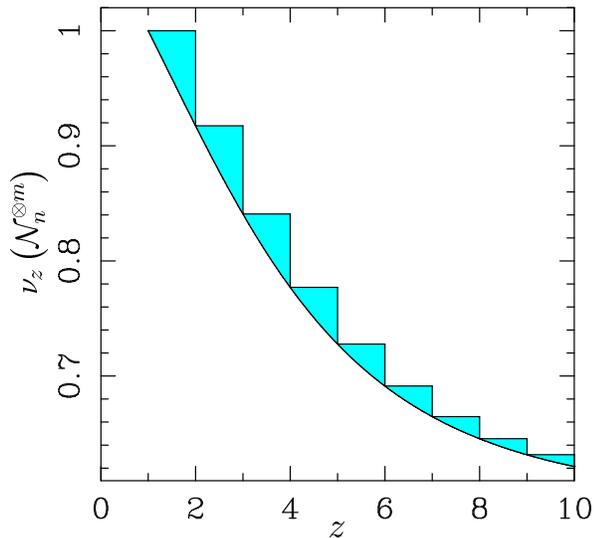}
\end{center}
\vspace{-.5cm}
\caption{Plot of the bounds of the
$z$-norm $\nu_{z}({\cal N}_n^{\otimes m})$ of the channel ${\cal
  N}_n$. The function $\nu_{z}({\cal N}_n^{\otimes m})$
 is restricted to the gray area which is limited
from above by the upper bound
of Eq.~(\ref{pro111}) and from below by the 
lower bound of Eq.~(\ref{pro112}).
The two curves meet for $z=k$ integer. Here $m=2$
  and $n=0.3$.}
\labell{f:ren}\end{figure}

The R\'enyi entropy of Eq.~(\ref{defrenyi}) is decreasing function of the
parameter ${z}$. As a matter of fact, it obeys the
inequality~\cite{zyc}
\begin{eqnarray}
\frac{{z}-1}{{z}}S_{{z}}(\rho)\geqslant
\frac{{z'}-1}{{z'}}S_{{z'}}(\rho)
\;\labell{pro},
\end{eqnarray}
which applies for any $z\geqslant z'$ and for all $\rho$.
This property and the monotonicity of
$S_z(\rho)$ respect to the norm of
Eq.~(\ref{norme}) can be used to derive the 
relation $\| \rho \|_z \leqslant \| \rho \|_{z'}$ which, when
applied to the output state of a channel ${\cal M}$,
implies
\begin{eqnarray}
\nu_{z}({\cal M}^{\otimes m}) \leqslant \nu_{z'}({\cal M}^{\otimes m})\;
\qquad z\geqslant z'\geqslant1 \;.\label{pro11}
\end{eqnarray}
In the case of the channel ${\cal N}_n$, by choosing 
$z'=k$ integer and using the identity (\ref{normach1}) we
obtain the upper bound for all $z\geqslant k$, i.e.
\begin{eqnarray}
\left[\frac{1}{(n+1)^k-n^k}\right]^{m/k}
\geqslant \nu_{z}({\cal N}_n^{\otimes m})\;.\label{pro111}
\end{eqnarray}
This bound must be compared with the lower bound 
\begin{eqnarray}
\left[\frac{1}{(n+1)^z-n^z}\right]^{m/z}
\leqslant \nu_{z}({\cal N}_n^{\otimes m})\;,\label{pro112}
\end{eqnarray}
for arbitrary $z$ 
that derives by considering as input of the $m$ successive
uses of the channel a tensor product of coherent states.
This two bounds are plotted in Fig.~\ref{f:ren}.

\section{Generalization}\label{s:altri}
In this section we show that the results obtained for
the channel ${\cal N}_n$ apply also to other Gaussian channel models.
The first group we analyze is formed by the  {\em classical} 
channels $\cal G$ 
where, as in the case of ${\cal N}_n$ 
the photonic signal from the sender is displaced 
randomly in phase space according to a Gaussian distribution.
The second group is formed by the channels $\cal L$ where
the input signal is linearly coupled to an external environment
prepared in a Gaussian state.
\subsection{Classical channels}\labell{s:classico}
Consider CP map $\cal G$ which transforms
the input state $\rho$ into the output state
\begin{eqnarray}
{\cal G}(\rho)&=&\int d^2 \zeta \;
\frac{\exp({-\zeta \cdot \Gamma \cdot \zeta^{\dag}}) }{\pi/
(2 \sqrt{\mbox{det}[\Gamma])}}\labell{dueG}\\
&\times& \exp\left[(a,a^\dag)\cdot\zeta^\dag
\right] \; \rho \;  \exp\left[- (a,a^\dag)\cdot\zeta^\dag \right],
\nonumber
\end{eqnarray}
where $\zeta=(\mu,-\mu^*)$ and 
\begin{eqnarray}
\Gamma&\equiv&\left[\begin{array}{lll}
u&&v^*\cr
v&&u
\end{array}
\right]
\;\labell{defGamma} \qquad u\geqslant |v|\;,
\end{eqnarray} 
is a $2\times2$ positive Hermitian matrix \cite{nota}.
For $\Gamma=\openone/(2n)$ the map 
${\cal G}$ gives ${\cal N}_n$ of Sec. \ref{s:channel},
while for generic $\Gamma$ the channel 
(\ref{dueG}) is the generalization of ${\cal N}_n$ 
to the case of non circularly symmetric 
distribution (\ref{tre}). 
As shown in App. {\ref{apA}}, the map $\cal G(\rho)$
can be decomposed according to the relation 
\begin{eqnarray}
{\cal G}(\rho)=
\Sigma^{\dag}(\xi)  \; 
{\cal N}_n( \Sigma(\xi)\; \rho\; \Sigma^{\dag}(\xi))
\; \Sigma(\xi) \,
\labell{rel}\;,
\end{eqnarray}
where
\begin{eqnarray}
n &=& 1/(2 \sqrt{u^2-|v|^2})\labell{enne}\\
\xi&=& \frac{v}{|v|} \mbox{arctanh}\left[\, \left(\frac{u-\sqrt{u^2-|v|^2}}
{u+\sqrt{u^2-|v|^2}}\right)^{1/2}\,\right] \; \labell{xi},
\end{eqnarray}
and where 
\begin{eqnarray}
\Sigma(\xi) \equiv \exp\left[ \left(\xi^* \; a^2-\xi\; (a^{\dag})^2
\right)/2 \right]
\labell{squeezing}
\end{eqnarray}
is the squeezing operator.
In other words, for any input state $\rho$,
 the output state  ${\cal G}(\rho)$
can be obtained by  applying  the squeezing operator
$\Sigma(\xi)$ to $\rho$, then 
sending it through the channel ${\cal N}_n$, 
and, finally, applying  the anti-squeezing 
transformation  $\Sigma(\xi)^{\dag}$. 
We can thus consider ${\cal N}_n$ as a simplified  
version of $\cal G$ where all the squeezing 
operations have been removed.

An important consequence of Eq.~(\ref{rel}) 
is that the $z$-norms of the channels ${\cal G}$ and 
${\cal N}_n$ are
identical. In fact, using the invariance of the norm (\ref{norme}) 
under the unitary operation $ \Sigma^{\dag}(\xi)^{\otimes m}$,
we can write the $z$-norm of $m$ uses of the channel $\cal G$ as
\begin{eqnarray}
\nu_z({\cal G}^{\otimes m}) &=& \max_{\rho\in {\cal H}^{\otimes m}} 
\|{\cal N}_n^{\otimes m}( \Sigma(\xi)^{\otimes m} \; 
\rho \;\Sigma^\dag(\xi)^{\otimes m}) \|_z \nonumber\\
&=& \max_{\rho\in {\cal H}^{\otimes m}} 
\|{\cal N}_n^{\otimes m}(\rho) \|_z \equiv \nu_z({\cal N}_n^{\otimes m})
\labell{normaG}
\end{eqnarray}
where, in the second identity, the unitary operator 
$\Sigma(\xi)^{\otimes m}$
has being incorporated in the definition of the input state $\rho$
of the $m$ uses of the channel. 
In particular, for $m=1$ and $z=k$ integer, 
Eqs.~(\ref{normaG}) and (\ref{normach1})  give the
$k$-norm for the single channel use of $\cal G$, i.e.
\begin{eqnarray}
\nu_{{k}}({\cal G}) =  
\left[\frac{(2\sqrt{u^2-|v|^2})^{k}}
{(1+2\sqrt{u^2-|u|^2})^k-1} \right]^{1/k}
\labell{normach2}\;.
\end{eqnarray}
According to the decomposition rule of Eq.~(\ref{rel}), such
a maximum is achieved for the anti-squeezed coherent states
\begin{eqnarray}
|\alpha;-\xi\rangle \equiv \Sigma^\dag(\xi) |\alpha\rangle.
\labell{anti}
\end{eqnarray}
In fact, feeding the channel $\cal G$ with this input is equivalent
(apart from an irrelevant unitary transformation) to feeding
${\cal N}_n$ with the coherent state $|\alpha\rangle$.
Moreover, for generic~$m$ Eq.~(\ref{normaG}) gives
\begin{eqnarray}
\nu_{{k}}({\cal G}^{\otimes m}) =  
\left[\frac{(2\sqrt{u^2-|v|^2})^{k}}
{(1+2\sqrt{u^2-|u|^2})^k-1} \right]^{m/k}
\labell{normach22}\;.
\end{eqnarray}
which proves the Amosov-Holevo-Werner conjecture
for the channel $\cal G$, at least for integer $z=k$.
As in the case of Eq.~(\ref{anti}), the input 
states that achieve the maximum (\ref{normach22}) can be 
obtained by anti-squeezing the states which achieve the
maximal output $z$-norm for the channel ${\cal N}_n$, i.e.  
$|\alpha_1;-\xi\rangle_1 \otimes
\cdots \otimes |\alpha_m;-\xi\rangle_m$.

\subsection{Linear-coupling channels}\labell{s:env}
The linear-coupling channel model 
represents a communication line where the input photons 
(described by the annihilation operator $a$) interact with 
an external environment (with annihilation operator $b$)
through the beam splitter unitary operator
\begin{eqnarray}
U=\exp\left[\,(a^\dag b-ab^\dag)\arctan\sqrt{\frac{1-\eta}\eta}\;\right]\;
\labell{defu},
\end{eqnarray}
which transforms the fields according to 
\begin{eqnarray}
a\longrightarrow U^\dag aU&=&\sqrt{\eta}\; a + \sqrt{1-\eta} \; b\nonumber\\
b\longrightarrow U^\dag bU&=&\sqrt{\eta}\; b - \sqrt{1-\eta} \; a \; ,
\labell{uno}
\end{eqnarray}
with $\eta\in[0,1]$ being the beam splitter
transmissivity.  For $\eta=1$, $U$ is the identity
and the input signal is decoupled from the environment; for
$\eta=0$, instead, $U$ is a swap operator which replaces
the input signal with the environment input state.
The CP map of the linear-coupling channel is obtained by coupling the 
input state of the signal $\rho$ with the input state of the
environment $\tau_b$ through $U$ and then by tracing away the 
mode $b$. The resulting output state is then
\begin{eqnarray}
{\cal L}(\rho)=\mbox{Tr}_b\left[U\:\rho\otimes\rho_b\:U^\dag\right]
\;\labell{mappa}.
\end{eqnarray}
If $\rho_b$ is a  Gaussian state, the CP map $\cal L$ is Gaussian.
In what follows we will assume $\rho_b$ to be the squeezed thermal
state 
\begin{eqnarray}
\rho_b= \Sigma_b^\dag(\xi) 
\; \tau_b(n) \; 
\Sigma_b(\xi)   
\labell{termico}
\end{eqnarray}
where $\Sigma_b$ and
$\tau_b(n)$ are, respectively, the squeezing operator and 
the thermal state (\ref{termico}) of the $b$ mode.
For the channel (\ref{mappa}) a decomposition 
rule analogous to Eq.~(\ref{rel}) applies, namely
(see App. {\ref{apB}})
\begin{eqnarray}
{\cal L}(\rho)=
\Sigma^{\dag}(\xi)  \; 
{\cal E}_n( \Sigma(\xi)\; \rho\; \Sigma^{\dag}(\xi))
\; \Sigma(\xi) \,
\labell{relE}\;,
\end{eqnarray}
with ${\cal E}_n(\rho)$ the CP map (\ref{mappa}) where
the environment is in the thermal state $\rho_b=\tau(n)$.
The connection between $\cal L$ and ${\cal E}_n$ is thus analogous
to the connection between $\cal G$ and ${\cal N}_n$.
In particular we can derive the following identity 
\begin{eqnarray}
\nu_z({\cal L}^{\otimes m}) = \nu_z\left({\cal E}_n^{\otimes m}\right)\;,
\labell{normaGNUOVA}
\end{eqnarray}
which applies for all $m$ integer and for all $z$.
Proving the Amosov-Holevo-Werner conjecture for ${\cal E}_n$ 
is equivalent to proving it for $\cal L$: moreover, the input
states which achieve the maximum output $z$-norm for $\cal L$
are obtained by anti-squeezing  the input states which achieve the
maximum for ${\cal E}_n$.

The channel ${\cal E}_n$ has been extensively 
studied in Ref.~\cite{entro} where it was shown that
it satisfies the relation
\begin{eqnarray}
{\cal E}_n(\rho)&=&\left({\cal N}_{(1-\eta)n}\circ{\cal
    E}_0\right) (\rho) \equiv
{\cal N}_{(1-\eta)n}\left( \;{\cal
    E}_0(\rho)\right)
\labell{pr3},
\end{eqnarray}
with ${\cal E}_0$ being the lossy map, where the input photons
interact with the vacuum state of the  environment.
Equation (\ref{pr3}) shows that the output of the channel ${\cal E}_n$
can be obtained first applying the lossy map to the input state $\rho$
and then feeding it into the classical channel ${\cal N}_n$. This
composition rule has two important consequences.  
On one hand, it implies
\begin{eqnarray}
\nu_z({\cal E}_n^{\otimes m}) =\nu_z
(({\cal N}_{(1-\eta)n}\circ{\cal
    E}_0)^{\otimes m})\leqslant \nu_z\left({\cal N}_{(1-\eta)n}^{\otimes m}\right).
\labell{normaGGG}
\end{eqnarray}
In fact, the maximization implicit in the second term is
performed on a set of input states which form a proper subset of the
input states which enter in the maximization of the third term.
On the other hand, since the lossy channel maps coherent input states
into coherent outputs according to the transformation \cite{c}
 \begin{eqnarray}
{\cal E}_0(|\alpha\rangle\langle 
\alpha|) = |\sqrt{\eta}\alpha\rangle \langle \sqrt{\eta}\alpha|
\label{lossy}\;,
\end{eqnarray}
equations (\ref{pr3}) and (\ref{normalpha}) 
show that when
${\cal E}_n$ and ${\cal N}_{(1-\eta)n}$ act on coherent inputs they produce the same output
\mbox{$z$-norm}. 
This is sufficient to prove that, at least for \mbox{$z=k$} integer, 
the inequality in Eq.~(\ref{normaGGG}) is replaced by an identity:
we have already established in fact that the maximum $k$-norm
of the channel ${\cal N}_n$ is achieved for a coherent state.
Hence we can establish the following identity
\begin{eqnarray}
\nu_{{k}}({\cal E}_n^{\otimes m}) =  \,  
\left\{\frac{1}{[(1-\eta)n+1]^k-[(1-\eta)n]^k}\right\}^{m/k}
\labell{normaGalla}\;,
\end{eqnarray}
which, analogously to Eq.~(\ref{normach22}),
shows that the Amosov-Holevo-Werner conjecture applies for the
channel ${\cal E}_n$ at least for all integer $k$. Moreover we know
that, as in the case of ${\cal N}_n$, tensor products of 
coherent states are sufficient to achieve the maximum
of Eq.~(\ref{normaGalla}).

\section{Conclusion}\labell{s:concl}
In this paper we have 
studied  various model of Gaussian bosonic channel (i.e.
the classical maps $\cal G$ of Sec. \ref{s:classico}
 and the linear coupling maps $\cal L$ of Sec. \ref{s:env})
and  we have shown that the  
Amosov-Holevo-Werner
conjecture (\ref{normach}) applies to them 
at least in the case of  $z=k$ integer.
In particular we have proved that
tensor products of squeezed coherent states 
are the inputs that achieve 
the maximum output \mbox{$k$-norm}
for the $m$ successive uses of these channels. 
In the case of the circularly symmetric channels 
${\cal N}_n$ and ${\cal E}_n$ the
optimal state are just tensor product of coherent
states.  
These properties imply that,
for all integer order greater than $2$,
the R\'enyi entropies at the output of the channels $\cal G$ and $\cal
L$ are additive, and
suggest that the same behavior should apply  also to all
the other orders (see Sec. \ref{s:other}).
In particular, it seems reasonable to believe that these
channel posses an additive Holevo information \cite{nostro,werner}.

\appendix

\section{Properties of the operator $\Theta$.}\labell{s:rengen}
In this appendix we prove that the operator $\Theta$ of Eq.~(\ref{re1})
has the tensor product structure of Eq.~(\ref{tensore})
and we derive the identity~(\ref{defqj}).

\subsection{Derivation of Eq.~(\ref{tensore}).}\labell{s:tensore}
By using the property
\begin{eqnarray}
D_s^\dag(\vec{\mu}) D_s(\vec{\nu}) = D_s(\vec{\nu}-\vec{\mu}) \,
\exp\left[\left(\vec{\nu}\cdot \vec{\mu}^\dag - \vec{\mu}\cdot
  \vec{\nu}^\dag\right)/2 \right]
\labell{identita}
\end{eqnarray}
of the multi-mode displacement operator defined in
Eq.~(\ref{multid}), 
the expression (\ref{re1}) of $\Theta$  yields
\begin{eqnarray}
\Theta
&\equiv&\int \frac{d^2\vi{\mu}}{(\pi n)^{mk}}
\:{e^{-\vi\mu\cdot
{\mathbb C}\cdot\vi\mu\,^\dag+\vi\mu\cdot
{\mathbb G}^\dag\cdot\vi{a}\,^\dag-\vi{a}\cdot {\mathbb G}
\cdot\vi\mu\,^\dag}}
\;\labell{reappa},
\end{eqnarray}
where we have introduced the complex linear vector 
\begin{eqnarray}
\vi{\mu}&\equiv& 
(\vec{\mu}_1;\;\vec{\mu}_2;\cdots;\;\vec{\mu}_k)\\
&=&(\mu_{11},\cdots,\mu_{1m};\;{\mu}_{21},\cdots,\mu_{2m};\;\mu_{k1},
\cdots,\mu_{km})\nonumber\;,
\end{eqnarray}
which has $km$ elements, and 
\begin{eqnarray}
\vi{a}&\equiv& 
(\vec{a}_1;\;\vec{a}_2;\cdots;\;\vec{a}_k)\\
&=&(a_{11},\cdots,a_{1m};\;{a}_{21},\cdots,a_{2m};\;a_{k1},
\cdots,a_{km})\nonumber
\end{eqnarray}
where
$a_{sr}$ is the annihilation operator associated with the $s$th copy of
the $r$th use of the channel.
In Eq.~(\ref{reappa}), $\mathbb{G}$ and 
\begin{eqnarray}
\mathbb{C}\equiv\frac\openone n+\frac{\mathbb{A}}{2} \;,
\label{defCappa}
\end{eqnarray}
 are now ${mk}\times{mk}$
real matrices ($\openone$ is the $mk\times mk$ identity), 
which are obtained, respectively, by tensoring to the $m$th power 
the matrices $G$ and $C$ of Eqs.~(\ref{defg}) and
(\ref{defC}). In particular
for $k\geqslant3$, $\mathbb G$ and $\mathbb A$ have the block form  
\begin{eqnarray}
\mathbb{G}&\equiv&G^{\otimes m}=\left[\begin{array}{rrrrrr}
-\openone&\openone&0&\cdots&0&0\cr
0&-\openone&\openone&\cdots&0&0\cr
0&0&-\openone&\cdots&0&0\cr
\vdots&&&\ddots&&\cr
0&0&0&\cdots&-\openone&\openone\cr
\openone&0&0&\cdots&0&-\openone\cr
\end{array}
\right]
\;\labell{defgappa}
\\
\mathbb{A}&\equiv&A^{\otimes m}=\left[\begin{array}{rrrrrr}
0&-\openone&0&\cdots&0&\openone\cr
\openone&0&-\openone&\cdots&0&0\cr
0&\openone&0&\cdots&0&0\cr
\vdots&&&\ddots&&\cr
0&0&0&\cdots&0&-\openone\cr
-\openone&0&0&\cdots&\openone&0\cr
\end{array}
\right]
\;\labell{defappa}
\end{eqnarray}
where now
$\openone$ and $0$ are the $m\times m$ identity
and null matrix respectively.
On one hand these equations 
show that the Gaussian in the integral~(\ref{reappa})
couples together all the annihilation operators $a_{sr}$
which have the same index $r$ (i.e. the operators
 $a_{1r}$,
$a_{2r}$, $\cdots$ and $a_{k,r}$ which enter in the definition of the
vector
$\vu{a}_r$ of Eq.~(\ref{ava})). On the other hand,
Eqs.~(\ref{defgappa}) and (\ref{defappa}) show that any two 
annihilation operators $a_{sr}$ and $a_{s'r'}$ with $r\neq r'$
are not coupled by the integral~(\ref{reappa}). 
We can hence write such an expression as a (tensor) product
of $m$ independent Gaussian integrals where only the modes associated
with $\vu{a}_r$ enters: by doing this and using the
cyclic symmetry of the matrices $\mathbb G$ and $\mathbb A$
we finally obtain Eq.~(\ref{tensore}).

\subsection{Derivation of Eq.~(\ref{defqj}).}\labell{s:ren}
The identity (\ref{defqj}) has been derived in 
\cite{entro1} by showing that the operator on  the left-hand-side
of this expression has the same symmetric characteristic function
of the operator $\Theta_{jr}$ of Eq.~(\ref{re2}). 
Here, instead, we show that these operator
coincide by calculating their matrix elements
in the Fock basis of the annihilation operator~$b_{jr}$.
For the sake of simplicity in the following we will omit the
indexes~$j$ and $r$.

Given $|p\rangle$ and $|q\rangle$
Fock states of $b$ with $p\geqslant q$, consider the quantity 
\begin{eqnarray}
\langle p |\Theta|q \rangle = 
\int d^2\nu  \frac{e^{-d|\nu|^2/|e|^2}}{\pi n|e|^2}
\:\langle p | D(-\nu)
|q \rangle
\;\labell{dix1}.
\end{eqnarray}
Following the derivation given in \cite{caves1} the
 matrix element in the integral can be expressed in 
terms of the Laguerre polynomials as 
\begin{eqnarray}
\langle p | D(-\nu)|q \rangle= \left(\frac{q !}{p !}\right)^{1/2}
 e^{-|\lambda|^2/2} \nu^{p-q} L_{q}^{(p-q)}(|\nu|^2).
\nonumber\\
\labell{dix2}
\end{eqnarray}
Replacing this expression in (\ref{dix1}) and using the
identities~\cite{grad}
\begin{eqnarray}
\int_0^\infty dx \; e^{-\gamma x} L^{(0)}_q(x) =
\frac{(\gamma -1)^q}{\gamma^{q+1}} \qquad  \mbox{Re}\gamma>0\;,
\labell{dix3}
\end{eqnarray}  
and 
\begin{eqnarray}
\int_0^{2\pi} d\varphi e^{i \varphi (p-q)}=2\pi \delta_{pq}
\labell{dix4}
\end{eqnarray}
where $\delta_{pq}$ is the Kronecker delta, 
we finally obtain
\begin{eqnarray}
\langle p |\Theta|q \rangle = \frac{2/n}{2d+|e|^2}
\left(\frac{2d-|e|^2}{2d+|e|^2}\right)^{p}
\labell{dix5}\; \delta_{pq} \;,
\end{eqnarray}
which proves the thesis.

\section{Decomposition of the maps $\cal G$ and $\cal L$}
\label{apA}
In this section we derive the decomposition rule of
Eqs.~(\ref{rel}) and (\ref{relE}) which allows to express
 the map
$\cal G$ of Eq.~(\ref{dueG}) in terms
of ${\cal N}_n$ and the map $\cal L$ of Eq.~(\ref{mappa}) 
in terms of ${\cal E}_n$, respectively.

\subsection{Derivation of Eq.~(\ref{rel})}
Consider the Hermitian matrix
\begin{eqnarray}
B&\equiv&\left[\begin{array}{lll}
\alpha&&\beta^*\cr
\beta&&\alpha
\end{array}
\right]
\;,\labell{defB}
\end{eqnarray}
with
\begin{eqnarray}
\alpha&=&\left(\frac{u+\sqrt{u^2-|v|^2}}{2\sqrt{u^2-|v|^2}}\right)^{1/2}
\nonumber\\
\beta&=&\frac{v}{|v|}\;\left(\frac{u-\sqrt{u^2-|v|^2}}{2\sqrt{u^2-|v|^2}}\right)^{1/2}
\label{defB1}\;,
\end{eqnarray}
where $u$ and $v$ are the elements of $\Gamma$ defined in Eq.~(\ref{tre}).
The matrix $B$ has determinant equal to $1$ and inverse
\begin{eqnarray}
B^{-1}&\equiv&\left[\begin{array}{rrr}
\alpha&&-\beta^*\cr
-\beta&&\alpha
\end{array}
\right]\;,\labell{defBinv}
\end{eqnarray}
which diagonalizes $\Gamma$ through the relation
\begin{eqnarray}
B^{-1}\cdot \Gamma \cdot B^{-1} &=&
\left[\begin{array}{ccc}
\sqrt{u^2-|v|^2}&&0\cr
0&&\sqrt{u^2-|v|^2}
\end{array}
\right].\labell{digonale}
\end{eqnarray}
Moreover, when applied to $(a,a^\dag)$ this matrix
produces the Bogoliubov transformation
\begin{eqnarray}
(c,c^\dag)&\equiv& (a,a^\dag)\cdot B^{-1}\nonumber\\
&=&\Sigma^\dag(\xi) \; (a,a^\dag) \; \Sigma(\xi)\;,
\label{bogo}
\end{eqnarray}
where $\Sigma(\xi)$ is  the squeezing operator defined in
Eq.~(\ref{squeezing}).
Using these properties we can obtain Eq.~(\ref{rel})
from Eq.~(\ref{dueG}) by performing a change  of integration
variables. In fact, for 
$\zeta\rightarrow \zeta \cdot B$ we have 
\begin{eqnarray}
{\cal G}(\rho)&=&\int d^2 \zeta \;
\frac{\exp[{-\zeta\cdot (B^{-1}\cdot \Gamma\cdot
B^{-1}) \cdot \zeta^{\dag}}] }{\pi/
(2 \sqrt{\mbox{det}[\Gamma]})}\nonumber\labell{dueapA}\\
&\times& \exp\left[(c,c^\dag)\cdot \zeta^\dag
\right] \; \rho \; \exp\left[- (c, c^\dag)\cdot
\zeta^\dag \right]\\
&=&  \int d^2 \mu \;\frac{\exp[-2\sqrt{u^2-|v|^2}|\mu|^2] }{\pi(2
\sqrt{u^2-|v|^2})} \nonumber\\
&\times& \Sigma^\dag(\xi) \; D(\mu) \;  \Sigma(\xi) \; \rho
\; \Sigma^\dag(\xi) \; D^\dag(\mu) 
\; \Sigma(\xi) \;,
\nonumber
\end{eqnarray}
which, according to Eq.~(\ref{tre}) and (\ref{enne})
coincides with the left-hand-side of Eq.~(\ref{dueG}).
\subsection{Derivation of Eq.~(\ref{relE}).}
\label{apB}
For the sake of clarity, in what follows
the operators which acts only on the environment will have the 
subscript $b$ while the
operators which act only on the input state will have the subscript $a$.
Using the relations (\ref{uno}) it is easy to show that 
the coupling operator $U$  
transforms  $a^2 + b^2$ into itself, i.e. that it
commutes with the operator $\Sigma_a(\xi)\Sigma_b(\xi)$ which
squeezes both the signal mode $a$ and the environment mode 
$b$ by the
same quantity $\xi$.  
Inserting the identity decomposition 
$\Sigma_a^\dag(\xi)\Sigma_a(\xi)=\openone$
in Eq.~(\ref{mappa}) and using the invariance of the
trace under cyclic permutation, 
the above property allows us to write Eq.~(\ref{mappa}) as
\begin{eqnarray}
{\cal L}(\rho) &=&  \Sigma_a^\dag(\xi) \;
\mbox{Tr}_b\left[U\,\left(\Sigma_a(\xi)\rho \Sigma_a^\dag(\xi)
\otimes\tau_b(n)\right)\,U^\dag\right] \; \Sigma_a(\xi) \nonumber\\
&=& \Sigma_a^\dag(\xi) \: {\cal E}_n ( \Sigma_a(\xi)\rho
\Sigma_a^\dag(\xi)) \:  \Sigma_a(\xi)
\;\labell{mappa1},
\end{eqnarray}
which proves the thesis (\ref{relE}).

 \end{document}